\newcounter{myctr}
\def\myitem{\refstepcounter{myctr}\bibfont\noindent\ifnum\themyctr>9\else\phantom{0}\fi\hangindent17pt\themyctr.\enskip}

\documentclass{ws-ijqi}

\usepackage{bbm}

\begin{document}



\title{Bracket states for communication protocols with coherent states}

\author{Alessia Allevi}

\address{Department of Science and High Technology and CNISM, Via Valleggio 11\\
Como I-22100 Italy\\
alessia.allevi@uninsubria.it}

\author{Stefano Olivares}

\address{Department of Physics, University of Milano and CNISM, Via Celoria 16\\
 Milano I-20133, Italy\\
stefano.olivares@fisica.unimi.it}

\author{Maria Bondani}

\address{Institute for Photonics and Nanotechnologies - CNR and CNISM, Via Valleggio 11\\
Como I-22100 Italy\\
maria.bondani@uninsubria.it}

\maketitle

\begin{history}
\received{Day Month Year}
\revised{Day Month Year}
\end{history}

\begin{abstract}
We present the generation and characterization of the class of bracket states, namely phase-sensitive mixtures of coherent states exhibiting symmetry properties in the phase-space description.
A bracket state can be seen as the statistical ensemble arriving at a receiver in a typical coherent-state-based communication channel. We show that when a bracket state is mixed at a beam splitter with a local oscillator, both the emerging beams exhibit a Fano factor larger than 1 and dependent on the relative phase between the input state and the local oscillator. We discuss the possibility to exploit this dependence to monitor the phase difference for the enhancement of the performances of a simple communication scheme based on direct detection. Our experimental setup involves linear optical elements and a pair of photon-number-resolving detectors operated in the mesoscopic photon-number domain.
\end{abstract}


\section{Introduction} \label{intro}	
Coherent states of light play a relevant role in practical communication protocols. One of the main advantages of these states over more exotic quantum states, such as the squeezed ones, is that they can propagate in free space and over long distances,\cite{agarwal10} only suffering attenuation and without altering their fundamental properties. It has also been demonstrated that such states can maximize the information transmitted in communication channels.\cite{giovannetti04} However, encoding information on multiple coherent states\cite{becerra13} requires the implementation of optimized strategies for their detection and discrimination, as they are non-hortogonal.\cite{chefles00}
During the last decade, many solutions, based on homodyne detection, or ON/OFF or photon-number resolving detection, have been theoretically\cite{ban97,takeoka03,olivares04,cariolaro10,assalini11} and experimentally investigated.\cite{cook07,wittmann10a,wittmann10b,muller12a,izumi12} The simplest setup is represented by a quasi-optimal discrimination scheme, in which the coherent states to be analyzed, namely  $|+ \alpha \rangle$ and $|- \alpha \rangle$, interfere with a local oscillator (LO) $|z \rangle$ at a high-transmissivity beam splitter (BS), whose outputs are measured by direct detection (Kennedy-like receiver\cite{kennedy}). Overall, the effect of the interference is to displace the input in order to obtain completely destructive or constructive interference at one of the outputs. The main limitations in the realization of such a system are, on the one hand, the existence of noise sources,\cite{olivares13} such as phase diffusion,\cite{genoni11} and, on the other one, the a-priori knowledge of the LO phase.\cite{muller12b}
Here we discuss the possibility to accomplish these two tasks by considering phase-sensitive mixtures of coherent states, which we will refer to as bracket states in view of their shape in the phase space. Our strategy is based on the use of linear optical elements and photon-number resolving detectors operated in the mesoscopic photon-number domain.

\section{The class of bracket states}
\begin{figure}
\vspace{-1.2cm} 
\centering
 \includegraphics[width=0.45\textwidth]{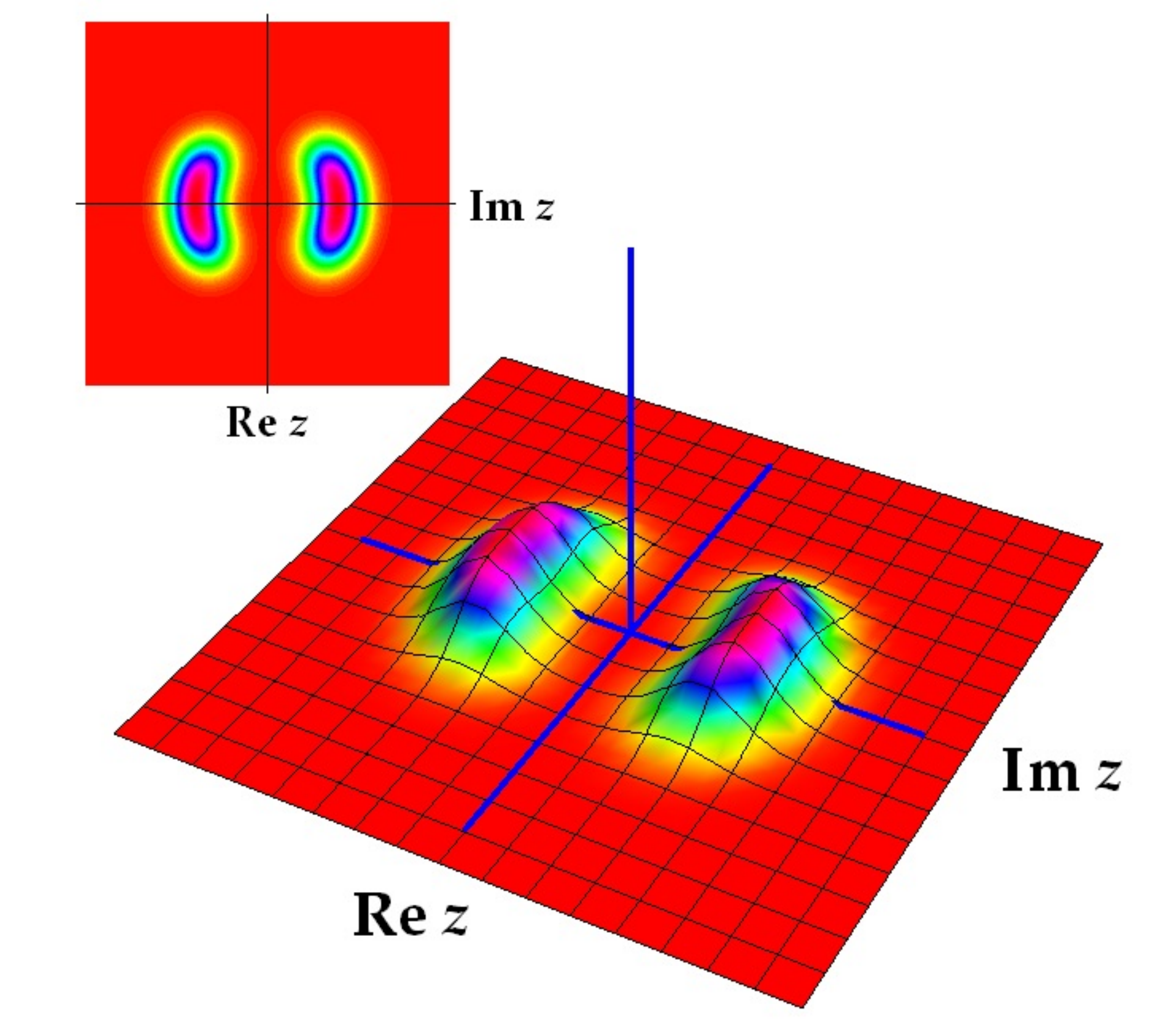}
 \vspace{-0.2cm} 
\caption{(Color online) Wigner function $W(z)$ of a bracket state with $\gamma=\pi/2$ and amplitude $b=2$. In the inset the contour plot is shown.}
\label{bracket}
\end{figure}

The bracket states are defined by the following density matrix
\begin{equation} \label{eq:bracket}
\varrho = \int_{-\gamma/2}^{+\gamma/2} \frac{d\psi}{\gamma} \frac{|be^{i\psi}\rangle \langle be^{i\psi}|+|-be^{i\psi}\rangle \langle -be^{i\psi}|}{2},
\end{equation}
with $\gamma \in [0,\pi]$ and without loss of generality we can assume $b \in  \mathbbm{R}$, $b \geq 0$. If $\gamma \rightarrow 0$, $\varrho$ reduces to the mixture of two coherent states, namely $|+b \rangle$ and $|-b \rangle$, thus representing the statistical ensemble of the states sent in binary communication channels with phase-shift-keyed (PSK) signals\cite{osaki96,kato99} and equal prior probability. The opposite case $\gamma = \pi$ corresponds
to having a phase-averaged coherent (PHAV) state\cite{OE12,JOSAB13} with amplitude $b$.  It is worth noting that PHAV states have been successfully used as decoy states to enhance the security of communication channels in key distribution protocols \cite{curty09}. It is important to notice that the parameter  $\gamma$ can be seen as the amplitude of an overall uniform phase noise affecting the generation and/or the propagation of the coherent signals.
For $\gamma < \pi$, the state $\varrho$ is phase-sensitive, as it is also evident in Fig.~\ref{bracket}, where the Wigner function 
\begin{equation} \label{Wigner}
W(z) = \frac{1}{\pi}\sum_{k=0,1}
\int_{-\gamma/2}^{+\gamma/2} \frac{d\psi}{\gamma}
\exp \left\{-2\, \left|z-(-1)^k\, b\, e^{i\psi}\right|^2 \right\},
\end{equation} 
of the bracket state with $\gamma = \pi/2$ and $b=2$ is shown.\\
\par
As a generic bracket state is essentially a balanced mixture of coherent states with the same energy but different phase, it has a Poisson photon-number statistics and Fano factor $F = 1$. 
\begin{figure}
\vspace{-0.7cm} 
\centering
\includegraphics[width=0.6\textwidth]{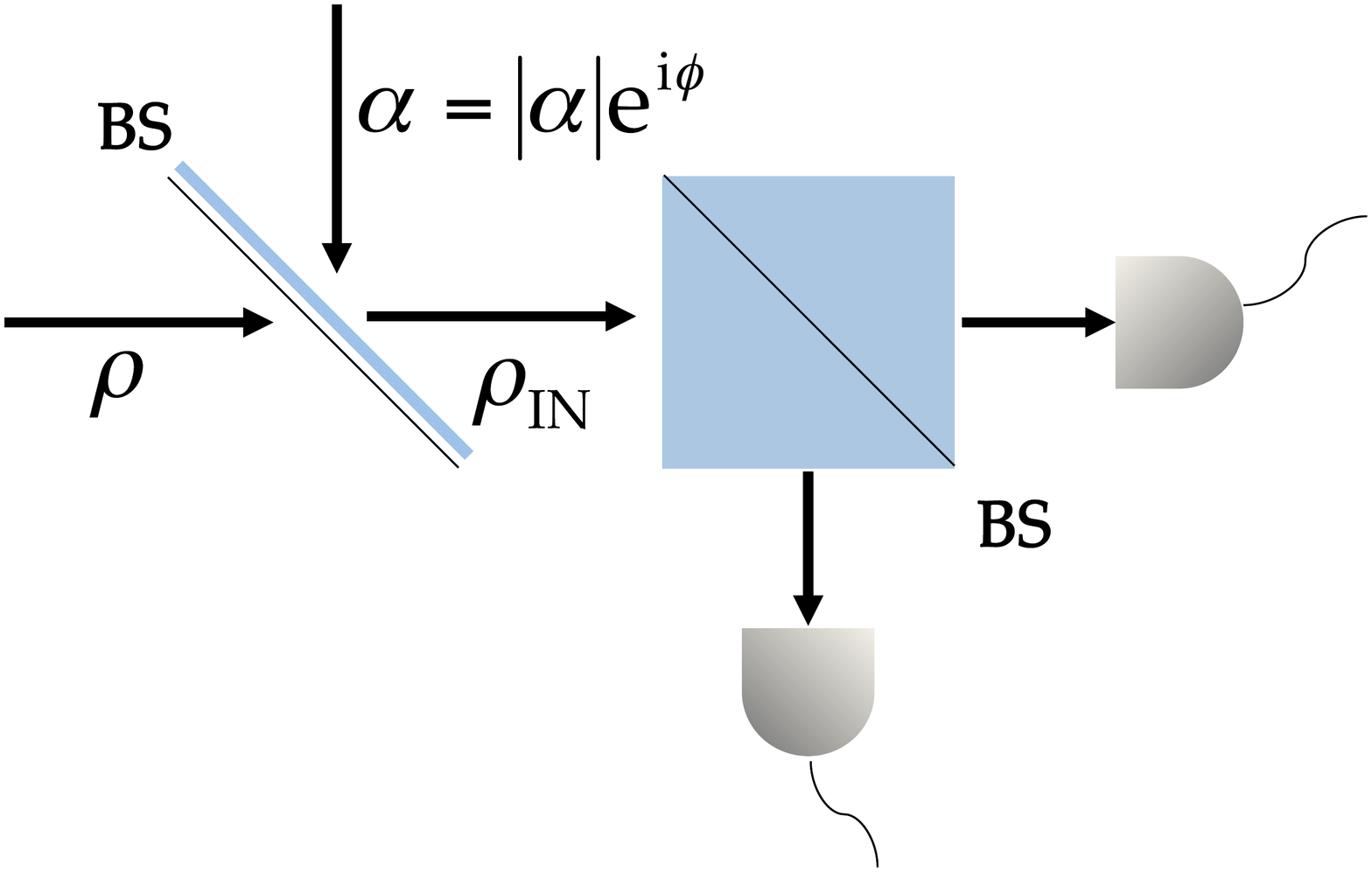}
\vspace{-0.7cm} 
\caption{(Color online) Scheme for the investigation of phase-dependent correlations
exhibited at a BS by a displaced phase-sensitive quantum state. See the text for details.}
\label{scheme} 
\end{figure}
When such a state is displaced by a coherent field $\alpha = |\alpha|e^{i \phi}$, which is the local oscillator, the first two moments of the photon-number distribution of the resulting state $\varrho_{\rm IN}$ in Fig.~\ref{scheme} become phase-dependent, namely:
\begin{equation} \label{meandispl}
\langle \hat{N} \rangle =
\langle \hat{N} \rangle_{\varrho} + |\alpha|^2 +
\sqrt{2}|\alpha| \langle \hat{x}_\phi \rangle_{\varrho}
\end{equation}
and
\begin{eqnarray} \label{vardispl}
{\rm Var} [\hat{N}] &=& {\rm Var}_{\varrho} [\hat{N}] + 2|\alpha|^2 {\rm Var}_{\varrho} \left[ \hat{x}_{\phi} \right],
\end{eqnarray}
where $\langle...\rangle_{\varrho} = {\rm Tr} \left[ \varrho...\right]$, ${\rm Var}_{\varrho} [\hat{N}] = \langle \hat{N}^2 \rangle_{\varrho} - \langle \hat{N} \rangle^2_{\varrho}$
and $\hat{x}_{\phi}$ is the quadrature operator
\begin{equation} \label{quadrature}
\hat{x}_{\phi} = \frac{\hat{a}^{\dag} e^{i\phi}+\hat{a} e^{-i\phi}}{\sqrt{2}},
\end{equation}
associated with the field mode $\hat{a}$, $[\hat{a},\hat{a}^{\dag}] = 1$.
The Fano factor of the displaced bracket state, which can be obtained from Eqs.(\ref{meandispl}) and (\ref{vardispl}), reads as follows 
\begin{equation} \label{fanobracket}
F = 
\frac{{\rm Var}[\hat{N}]}{\langle \hat{N} \rangle}
= \frac{b^2 + 2|\alpha|^2 {\rm Var}_{\varrho}[\hat{x}_{\phi}]}{b^2 + |\alpha|^2} \ge 1,
\end{equation}
where 
\begin{equation}
{\rm Var}_{\varrho} [\hat{x}_{\phi}] =  Tr[\varrho(\hat{x}_{\phi}- \langle \hat{x}_{\phi} \rangle)^2] =
\frac12 + b^2 \left[ 1+ \cos(2 \phi) \frac{\sin \gamma}{\gamma} \right] \label{variance}
\end{equation}
is the variance of the quadrature operator, in which we used $\langle \hat{x}_\phi \rangle = 0$, $\forall \phi$.
\par
When the displaced bracket state $\varrho_{\rm IN}$ is sent through a BS with transmissivity $\tau$, as shown in Fig.~\ref{scheme}, the two output beams may exhibit intensity correlations.\cite{OLcorr} In this particular case, the intensity correlation coefficient $\Gamma$ between the output beams can be written as a function of the Fano factor of $\varrho_{\rm IN}$ as follows:
\begin{equation} \label{eq:corr}
\Gamma(\phi) = \frac{[F(\phi)-1]\sqrt{\tau (1 - \tau)}}{\sqrt{[F(\phi) \tau + (1 - \tau)][F(\phi)(1 - \tau) + \tau]}},
\end{equation}
that reduces to
\begin{equation}
\Gamma(\phi)=
\frac{F(\phi)-1}{F(\phi)+1}. \label{eq:50corr} 
\end{equation}
for a balanced BS ($\tau = 1/2$). Therefore, if $F>1$, intensity correlations arise between the two emerging beams.
\par
In the following, we investigate the experimental behavior of $F(\phi)$ and $\Gamma(\phi)$ as functions of the phase of the displacement amplitude.

\section{Experimental results and discussion}
\begin{figure}
\centering
\includegraphics[width=0.45\textwidth]{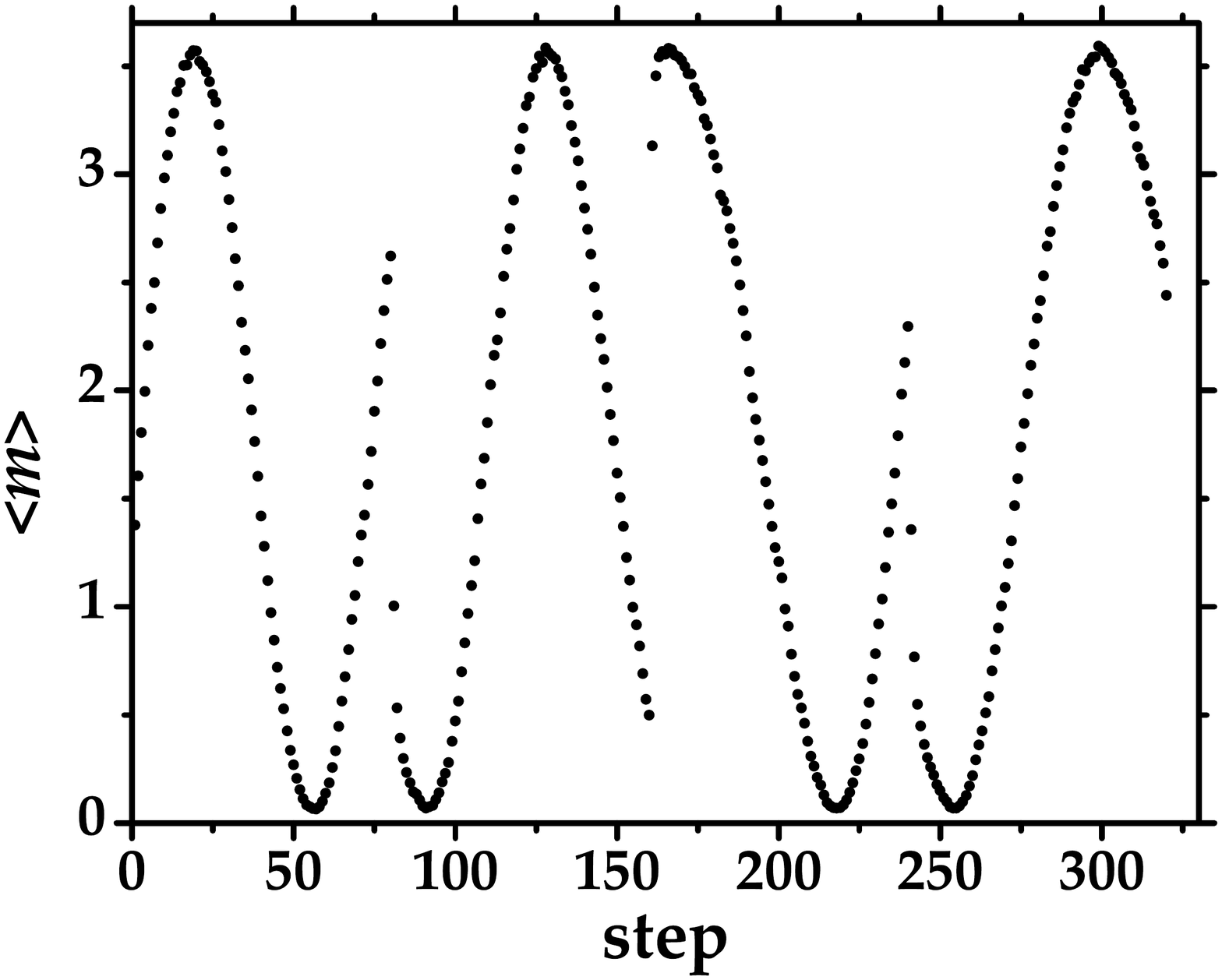}
\includegraphics[width=0.45\textwidth]{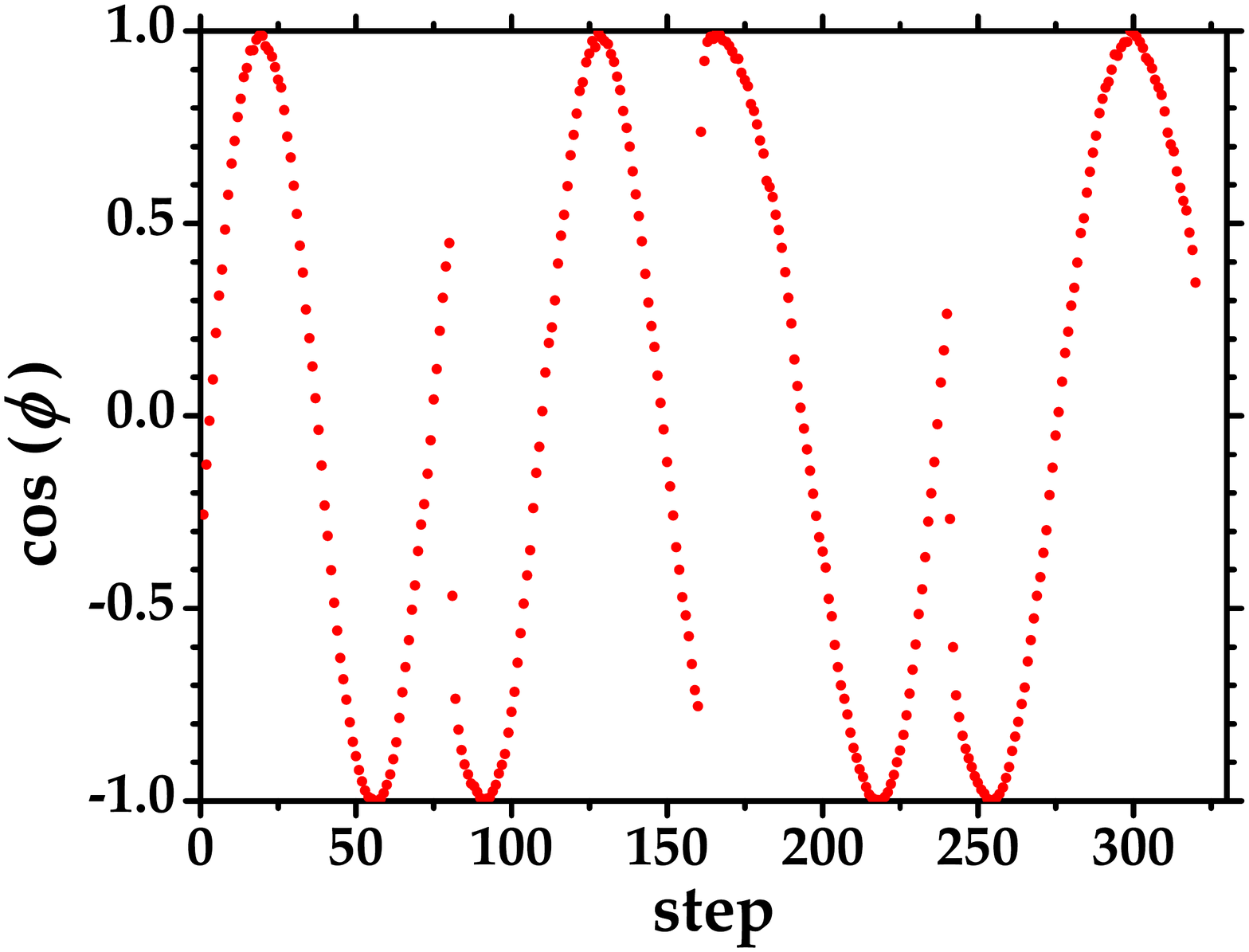}
\includegraphics[width=0.45\textwidth]{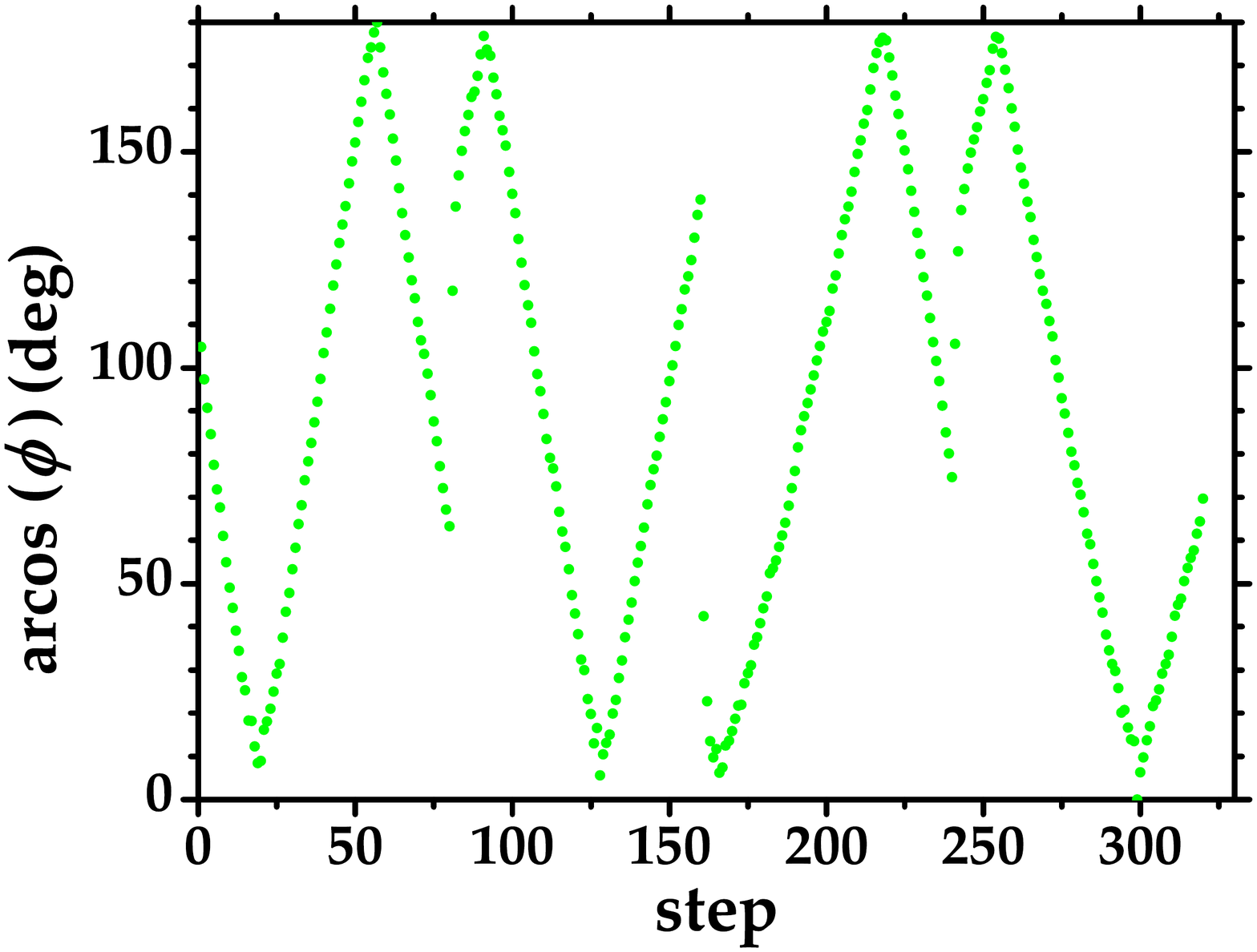}
\includegraphics[width=0.45\textwidth]{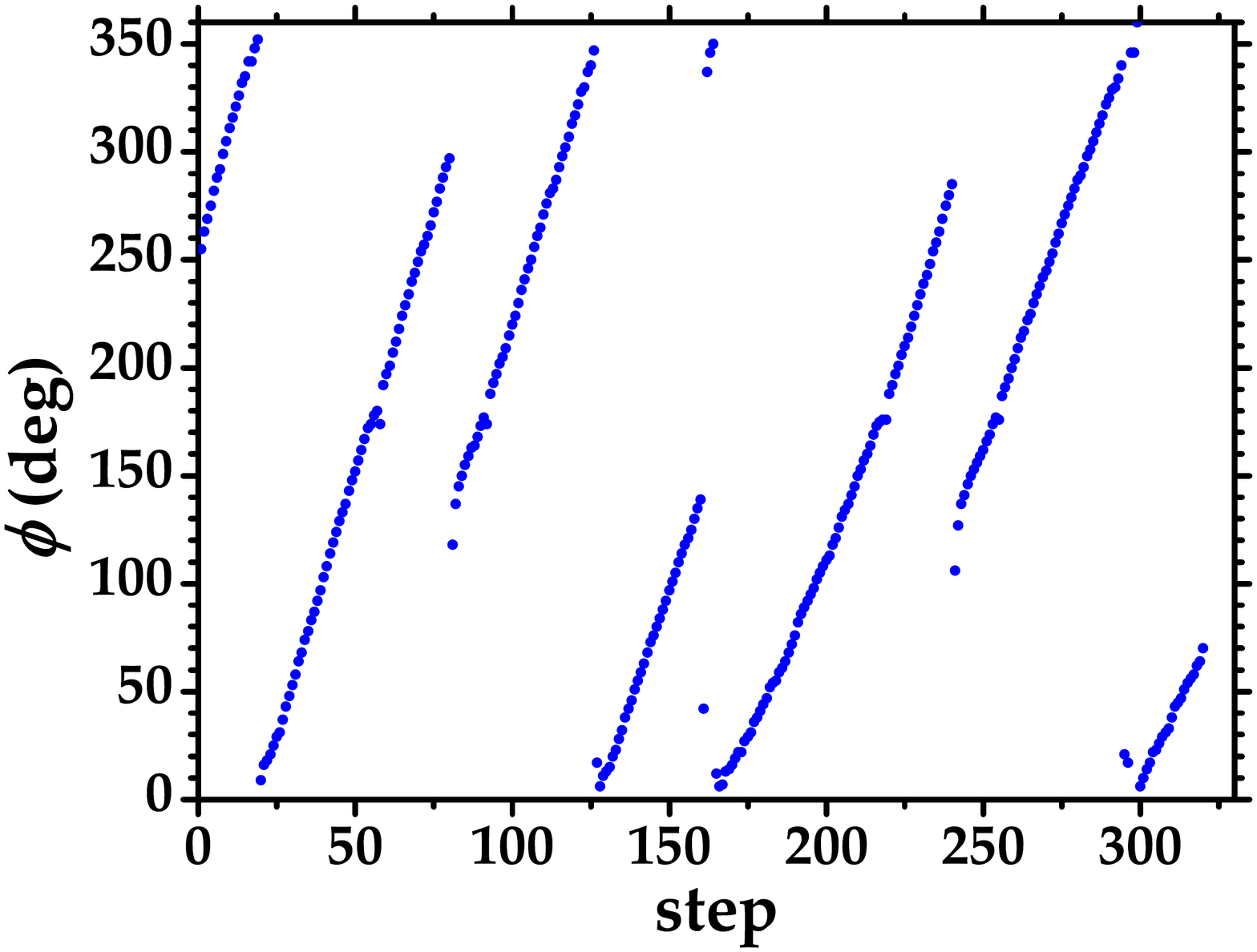}
\vspace{-0.4cm} 
\caption{(Color online) Upper panels: Mean number of photons detected by one of the two HPDs (left) and experimental cosine
of the interference pattern (right); lower panels: experimental arcosine (left) and relative phase values (right).
All the quantities are plotted as functions of the step of piezoelectric movement.}
\label{phase}
\vspace{-0.4cm} 
\end{figure}

\begin{figure}
\centering
\includegraphics[width=0.45\textwidth]{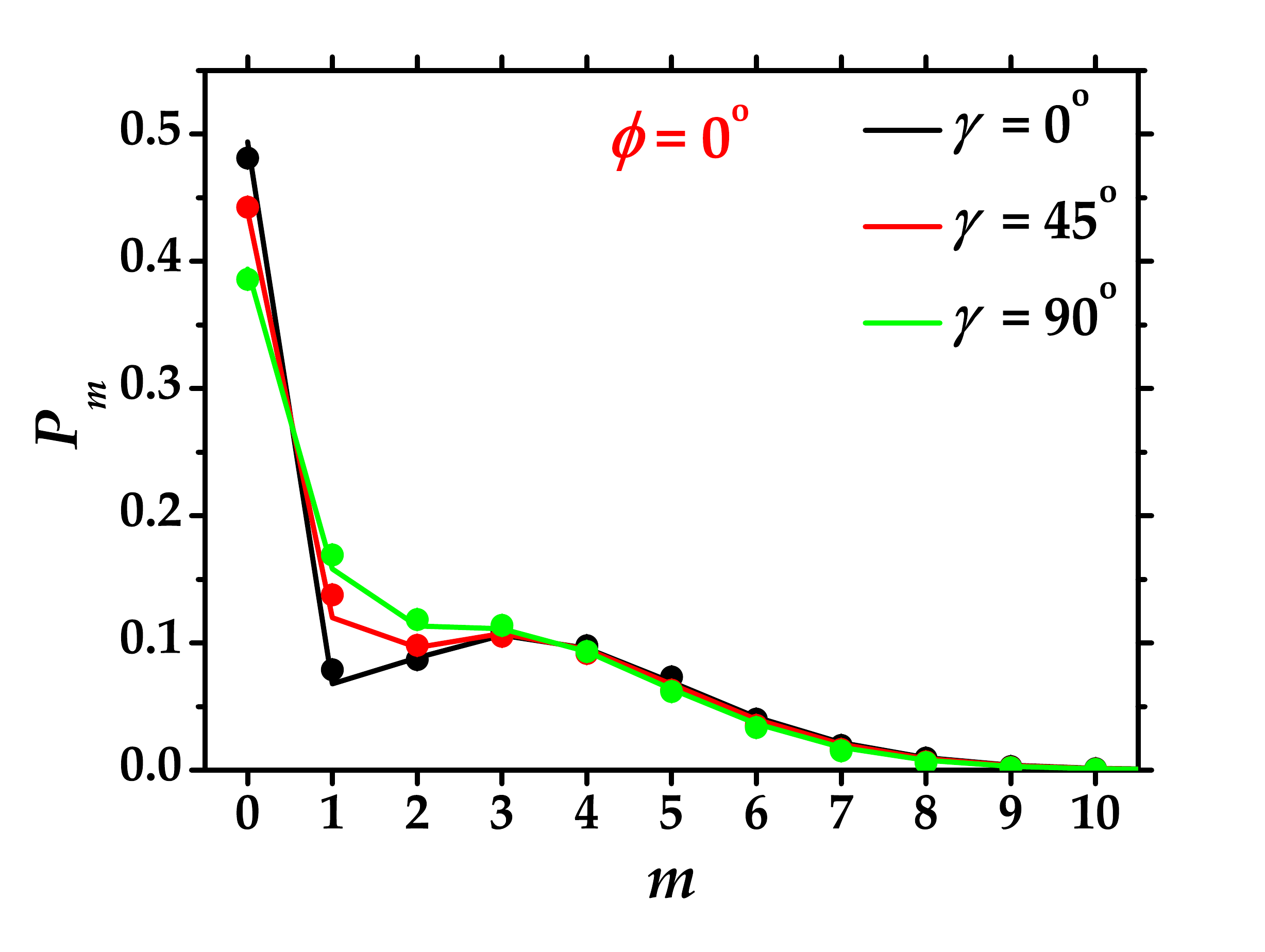}
\includegraphics[width=0.45\textwidth]{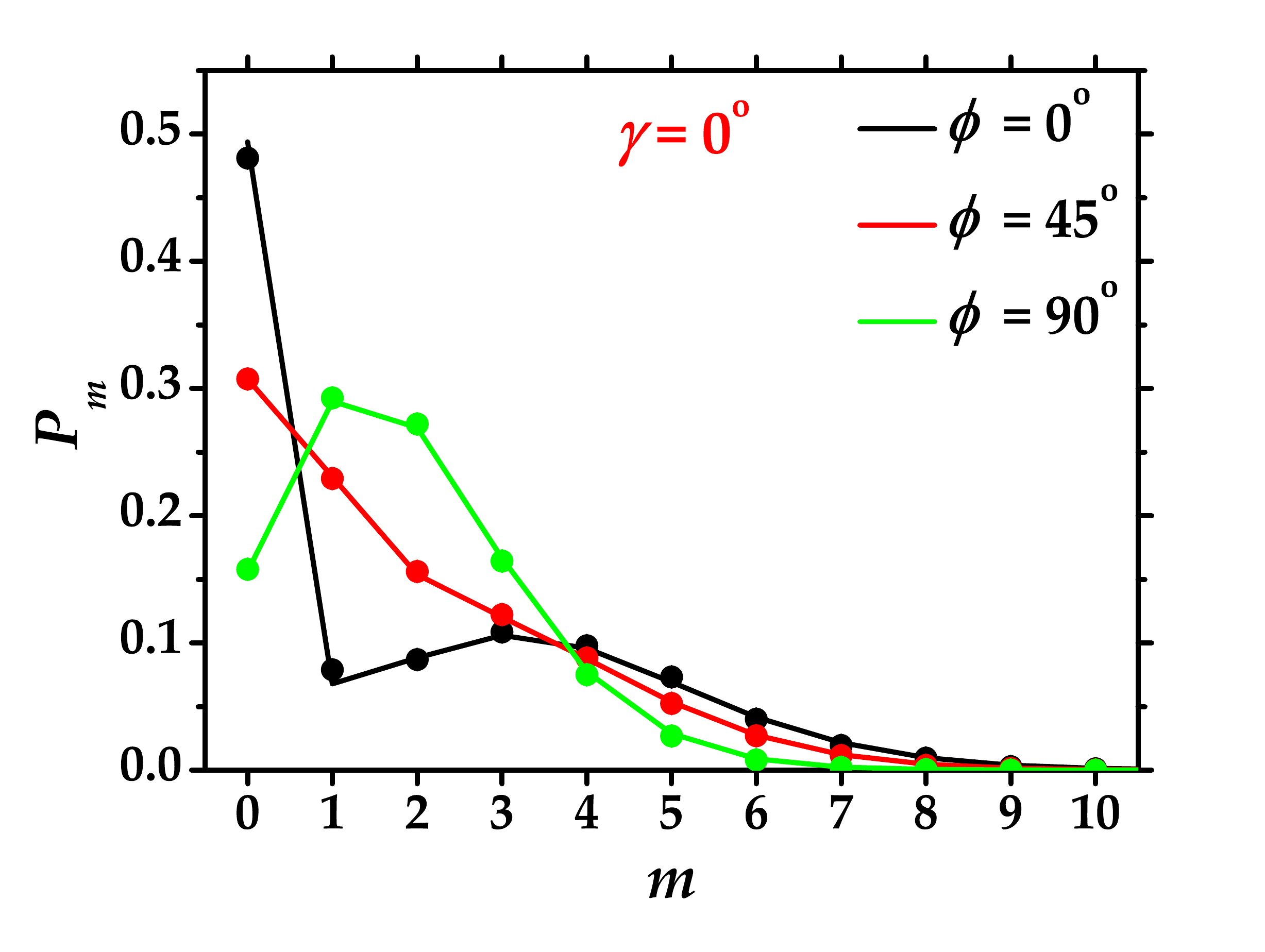}
\includegraphics[width=0.45\textwidth]{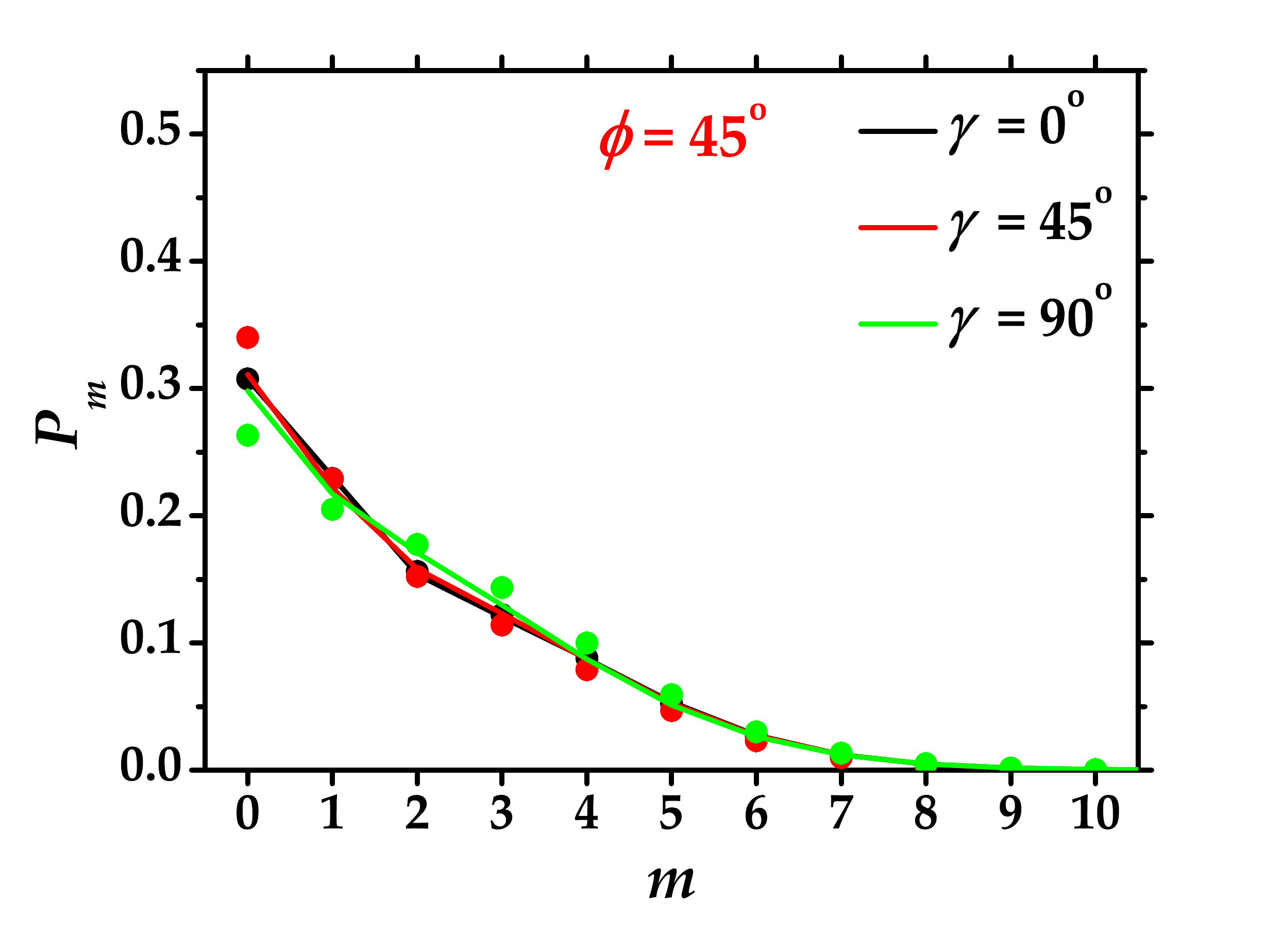}
\includegraphics[width=0.45\textwidth]{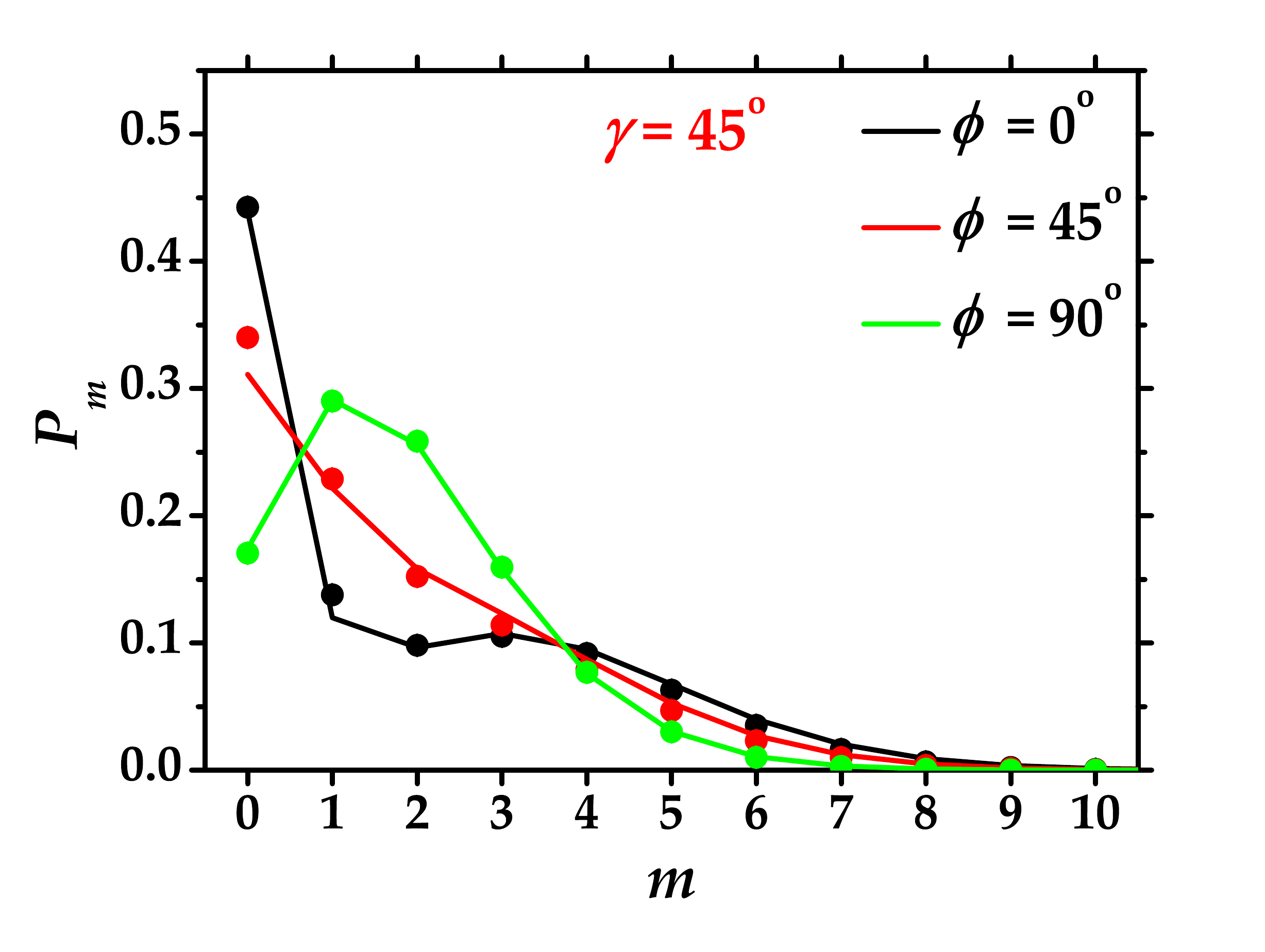}
\includegraphics[width=0.45\textwidth]{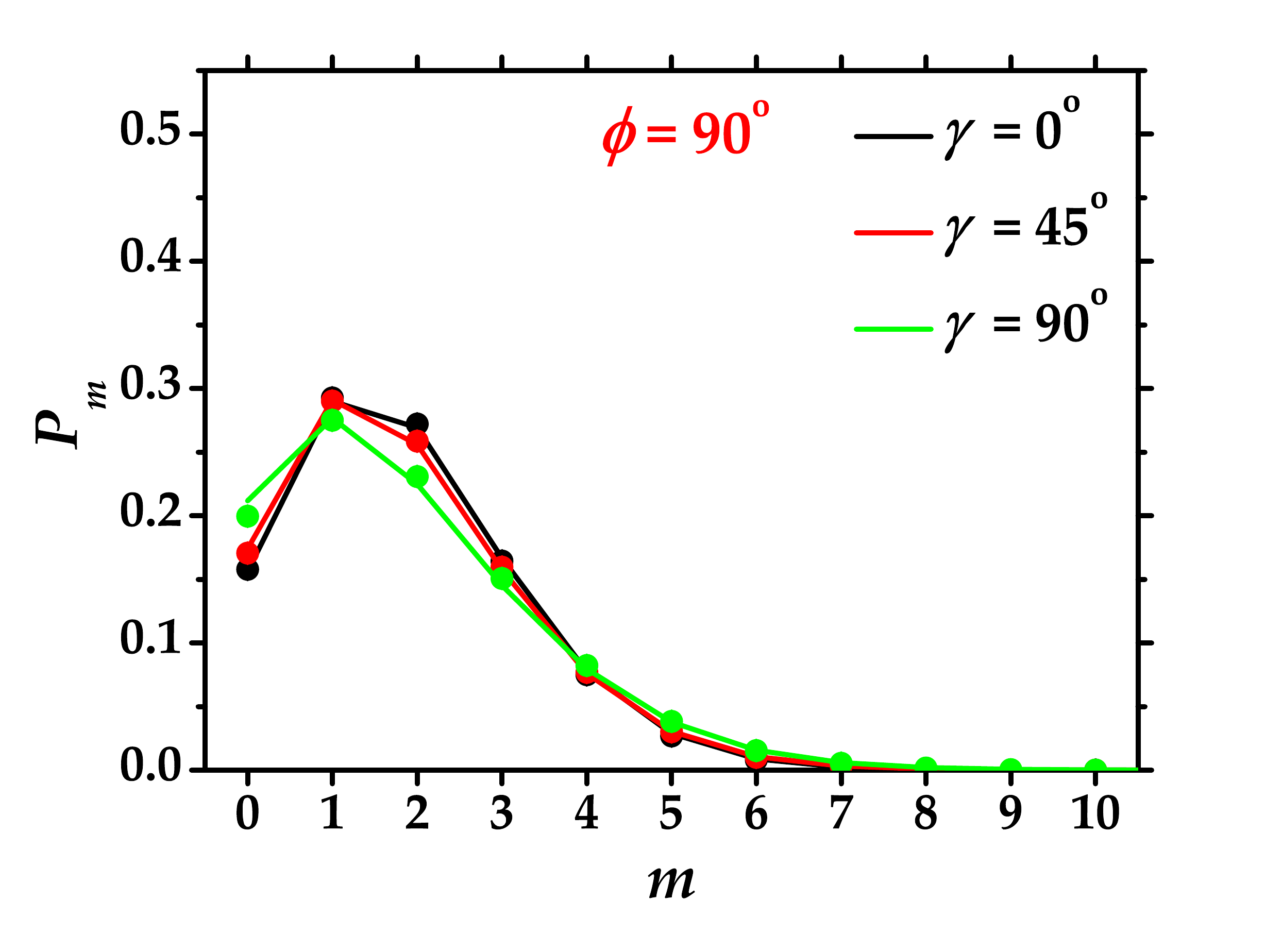}
\includegraphics[width=0.45\textwidth]{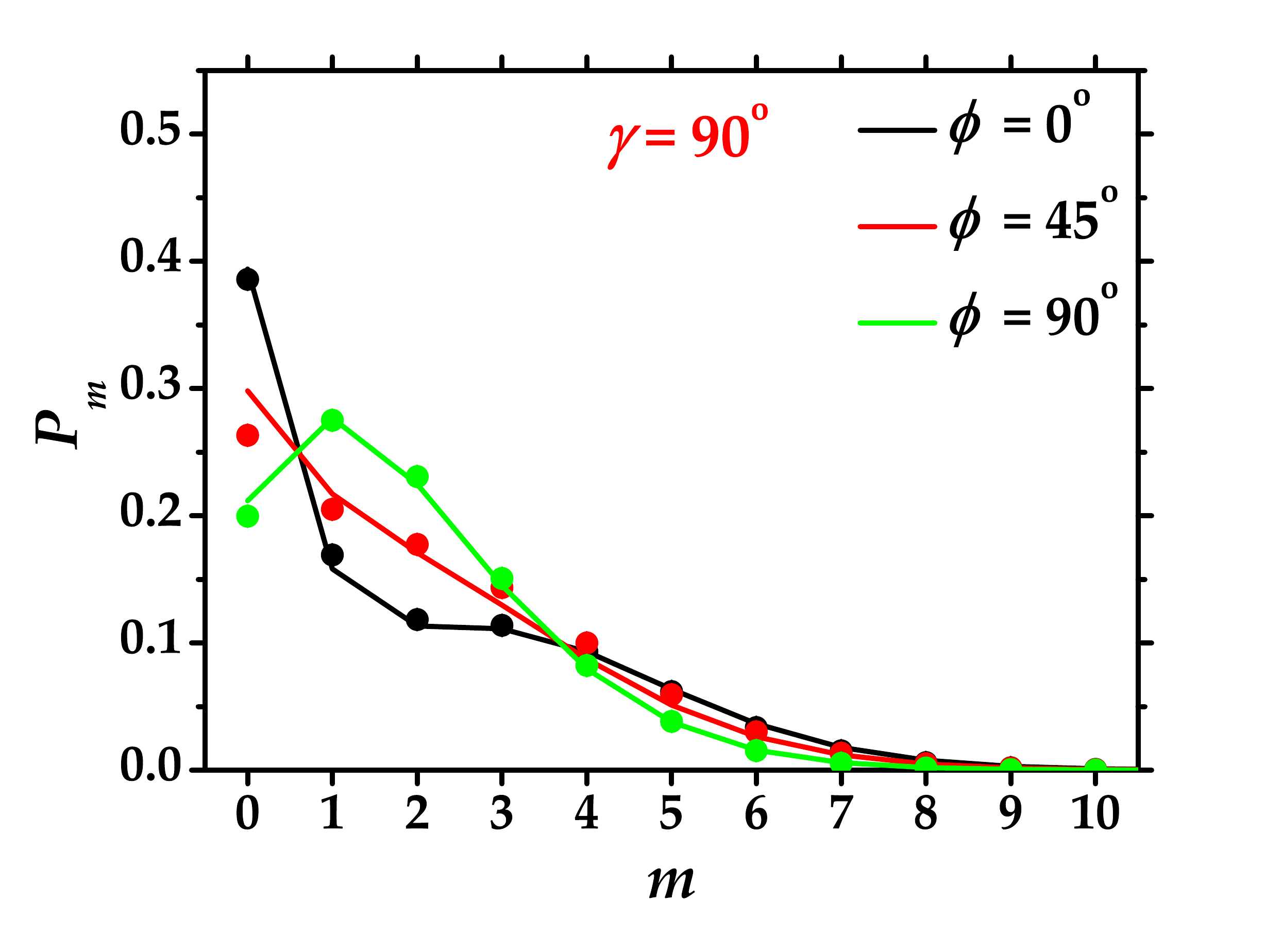}
\vspace{-0.2cm} 
\caption{(Color online) Detected-photon distributions of displaced bracket states for different choices of the relative phase $\phi$ (left panels) and of the interval $\gamma$ (right panels). 
In each panel the experimental data are shown as colored dots and the theoretical predictions as colored lines according to the same choice of colors.}
\label{statistics}
\vspace{-0.2cm} 
\end{figure}

\begin{figure}
\centering
\includegraphics[width=0.45\textwidth]{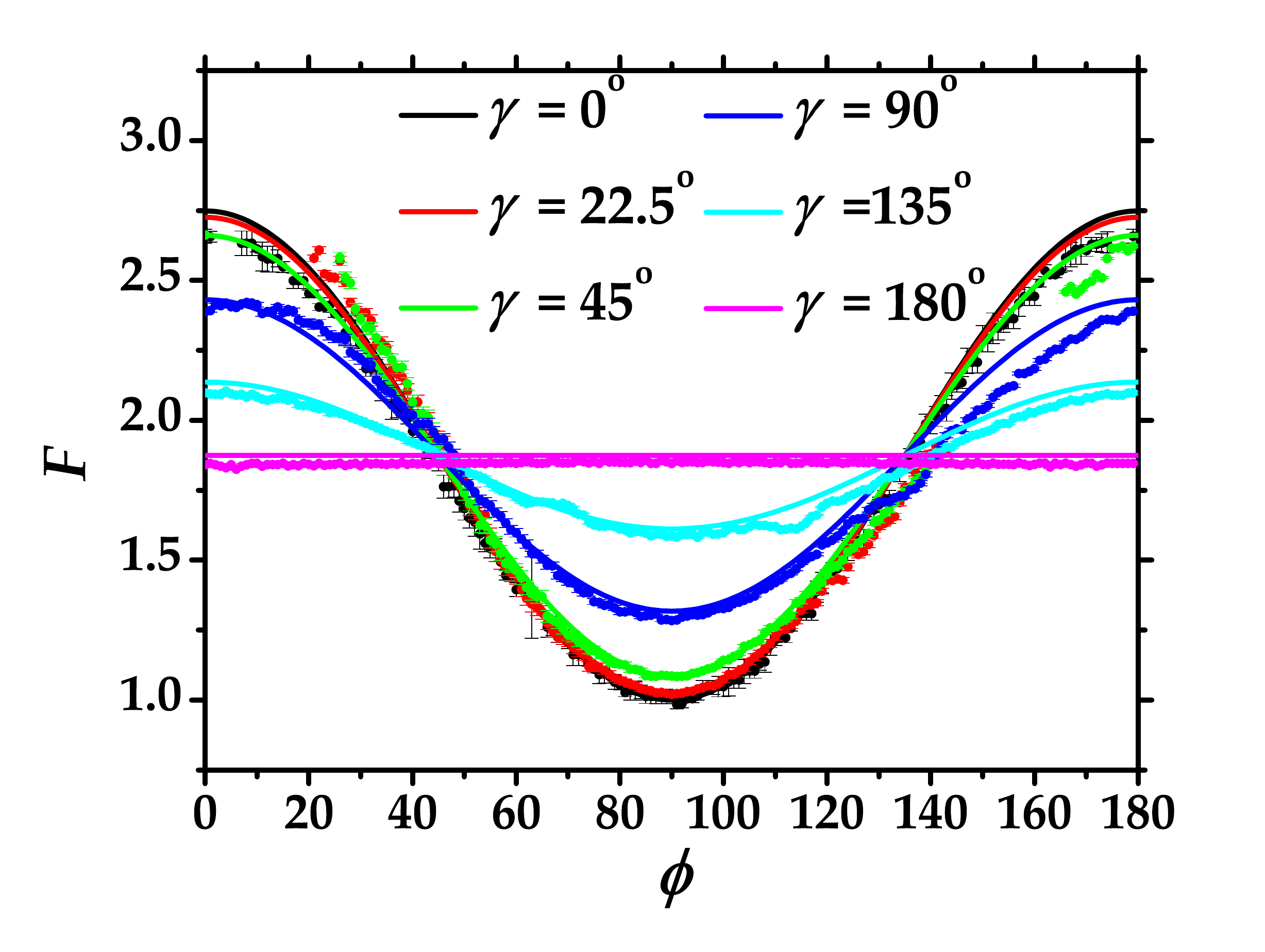}
\includegraphics[width=0.45\textwidth]{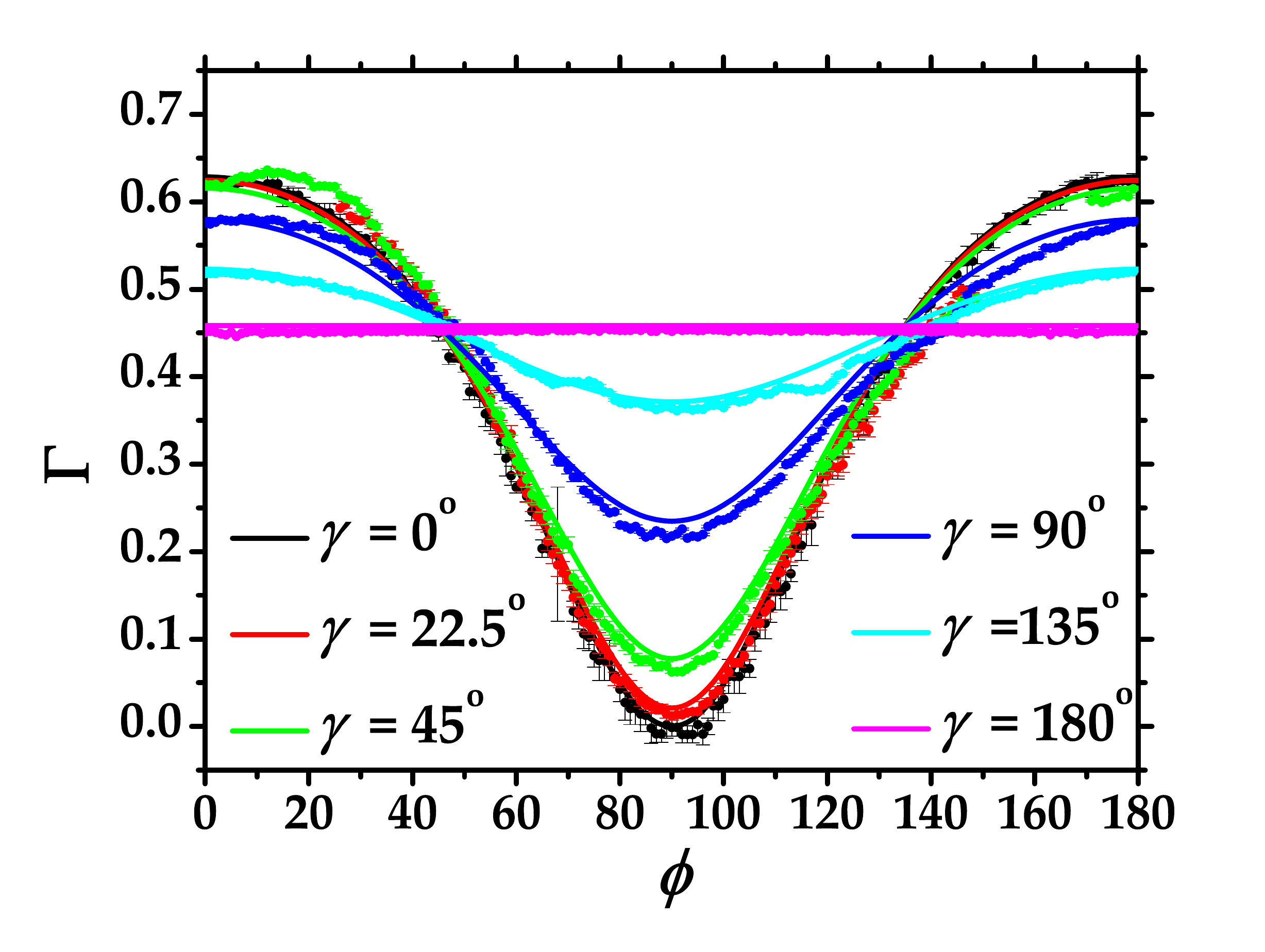}
\vspace{-0.2cm} 
\caption{(Color online) Fano factor for detected photons in one of the two detection arms (left panel) and intensity correlation coefficient (right panel) as functions of the relative phase $\phi$ for different choices of the interval $\gamma$.
Colored dots: experimental data, colored lines: corresponding theoretical predictions calculated in the actual experimental values.}
\label{fig:fano}
\vspace{-0.2cm} 
\end{figure}

The experimental generation of bracket states was obtained by exploiting the second-harmonics (523-nm wavelength, 5-ps pulse duration) of a 
mode-locked Nd:YLF laser amplified at 500 Hz (High-Q Laser Production). 
\begin{figure}
\vspace{-1.2cm} 
\centering\includegraphics[width=0.6\textwidth]{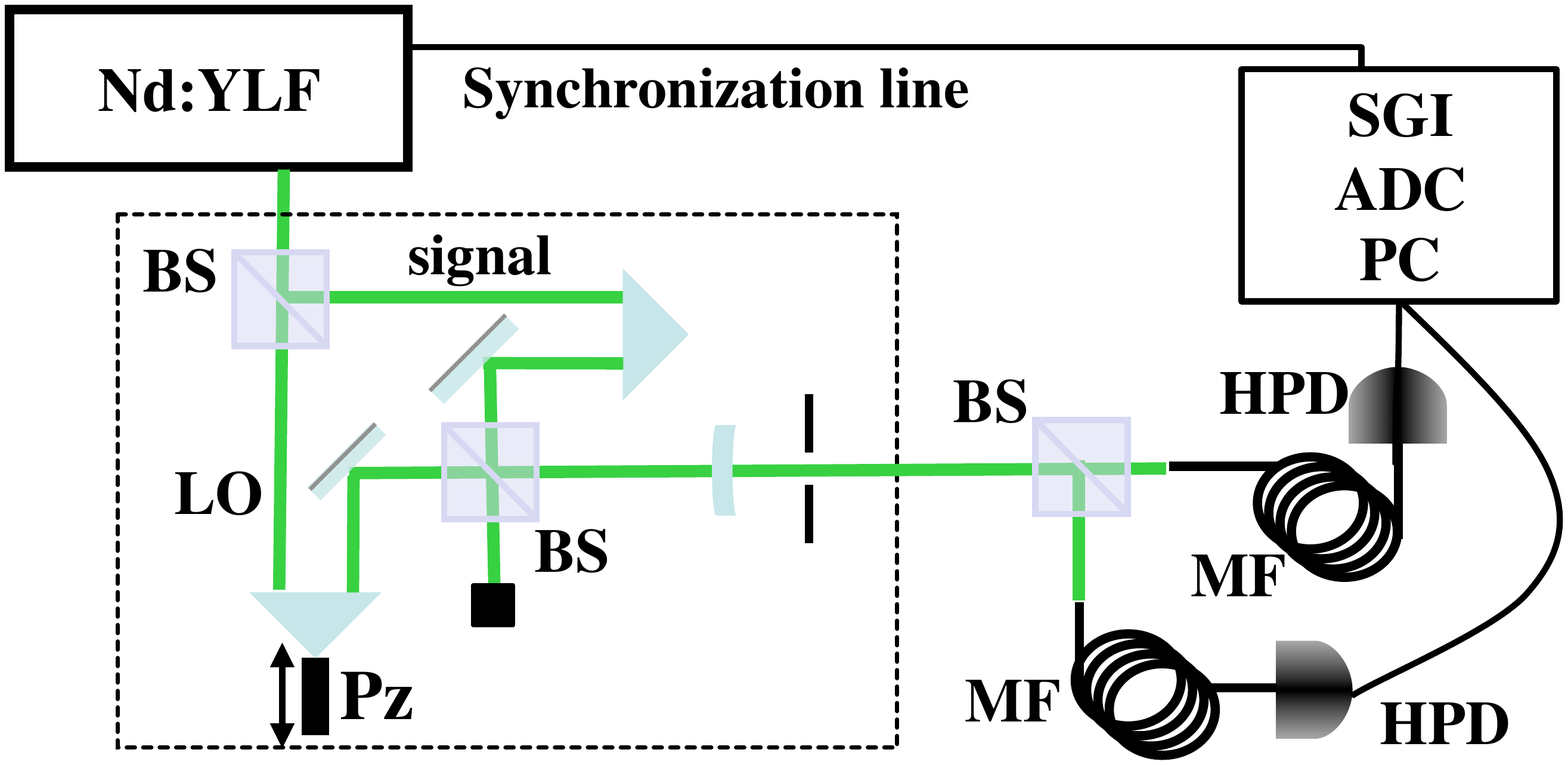}
\vspace{-0.7cm} 
\caption{(Color online) Sketch of the experimental setup. The laser beam is sent to a Mach-Zehnder interferometer, in which the relative phase
between the local oscillator (LO) and the signal is changed by means of a piezoelectric movement (Pz). One output of the interferometer is
suitably selected and divided at a BS, whose two outputs are delivered to two hybrid photodetectors (HPD) by means of two multimode fibers (MF).
The amplified output of each detector is synchronously integrated (SGI), digitized (ADC) and processed offline (PC).}
\label{setup}
\vspace{-0.2cm} 
\end{figure}
According to the experimental setup shown in Fig.~\ref{setup}, the linearly-polarized pulses were sent to a Mach-Zehnder interferometer: 
one of its two mirrors was mounted on a piezoelectric movement, whose displacement was 
operated step-by-step in order to change the relative phase between the two arms. In particular, we considered 320 different values of phase $\phi$. 
The displaced state we obtained in such a way was then sent to a further BS, whose
outputs were collected by two multimode fibers and delivered to a pair of hybrid photodetectors (HPD, R10467U-40, maximum quantum efficiency $\sim$ 0.5
at 500~nm, 1.4-ns response time, Hamamatsu). 
The output of each detector was amplified (preamplifier A250 plus amplifier A275, Amptek), synchronously integrated (SGI, SR250, Stanford) 
and digitized (AT-MIO-16E-1, National Instruments). As already explained in Refs.~[\refcite{OLcorr,ASL}], such a detection apparatus allows us
to reconstruct not only the statistics of detected photons but also to retrieve the shot-by-shot intensity correlations. 
Moreover, by following the procedure presented in~[\refcite{JosaBwigner}], we are also able to determine the actual value of the phase $\phi$ at
each piezo position, independent of the regularity and reproducibility of the movement. The method is simply based on the
linearity of our detectors. In fact, by monitoring the mean number of detected photons as a function of the piezolelectric movement, an interference pattern emerges. The normalization of such a behavior between $-1$ and $+1$ allows us to fit the experimental data with a cosine function and, thus, to directly estimate the value of $\phi$. The strategy we followed is well represented in Fig.~\ref{phase}, where we plot the different steps of 
our procedure.
Thanks to this phase determination, the bracket states were obtained in post-selection by combining a set of data corresponding to an interval $\gamma$ around $\phi$ 
and appending it to a second set corresponding to an interval with the same amplitude but with opposite phase.
It is interesting to notice that for different choices of $\gamma$ and $\phi$, the phase-sensitive nature of bracket states is evident not only in the behavior of the Fano
factor and of the correlation coefficient as anticipated in the previous Section, but also in the statistics of detected photons.
For instance, in the six panels of Fig.~\ref{statistics} we plot the detected-photon distributions of bracket states for different values of $\gamma$ and $\phi$.
In each panel of the figure we show the corresponding theoretical statistics for detected photons, which is obtained by numerically
integrating the trace of the displaced bracket state and by taking into account that, being this kind of state classical, the functional form is invariant under Bernoullian distribution.\cite{intech}      
The good agreement between experimental data and theory can be quantified by calculating the fidelity
$F = \sum_{m=0}^{\bar{m}} \sqrt{P_{\rm th}(m)P(m)}$, in which $P_{\rm th}(m)$ and $P(m)$ are the theoretical and
experimental distributions, respectively, and the sum is extended up to the maximum detected photon number $\bar{m}$ above which both
$P_{\rm th}(m)$ and $P(m)$ become negligible, that is $P(m) < 10^{-7}$ for $m>\bar{m}$. For all the statistics presented in Fig.~\ref{statistics} we obtained very high values of $F$ ($\ge 0.999$).\\
\indent
The experimental behavior of the Fano factor and of the corresponding intensity correlation coefficient as functions of the relative phase $\phi$
are shown in Figs.~\ref{fig:fano}.
In both figures we plot the experimental data (colored symbols) and the corresponding theoretical expectations (colored lines) for different choices of
$\gamma$ values. 
It is worth noting that $F$ and $\Gamma$ are periodic functions of the angle $\phi$ for all values of $\gamma$ in the interval $[0, \pi)$.
For $\gamma = \pi$ we can instead recognize the independence of the phase-averaged coherent state from the relative phase in the straight horizontal line.
In principle, the capability of our detection apparatus to reveal the dependence of the output on the relative phase $\phi$ makes our scheme particularly interesting
for the implementation of PSK communication protocols, as it avoids both any \emph{a-priori} knowledge of the phase and any preliminary communication between the sender and the receiver.
The dependence of $F$ on the relative phase $\phi$ allows us to satisfy this requirement by simply investigating the statistical properties of the displaced bracket state. Of course, in the accomplishment of this task, the use of detectors able to clearly discriminate photons plays a key role.
Moreover, the class of bracket states can be used to investigate the effect of phase noise in quantum state discrimination protocols by simply changing the value of the variables $\phi$ and $\gamma$ in the calculation of the error probability. In fact, by setting the integration interval equal to 0, we can produce and characterize coherent states with any phase between 0 and $\pi$, thus having the possibility to experimentally investigate the effect of a dephasing in the preparation of the states $|+b \rangle$ and $|-b \rangle$. With our scheme we can also simulate a phase-diffusion-like effect by setting the central phases equal to 0 or $\pi$ and choosing values of $\gamma$ different from 0.   
\par
As a matter of fact, here we have just discussed the {\it capability} of our apparatus to reveal the phase dependence, and, therefore, to monitor phase differences between the input state and the local oscillator. However, it is worth noting that, in general, the estimation of the phase difference by inversion methods may be not the best strategy and a Bayesian analysis could be needed or preferred.\cite{oli:bayes:HD,BT:bayes:HD,geno:exp:HD}

\section{Conclusions and future perspectives}
In conclusion, we have presented the generation and characterization of the class of bracket states. The excellent agreement between the experimental data and the theoretical prediction suggests the possibility to use our experimental scheme for the investigation of communication protocols in the presence of phase noise.
The linearity of our detection system, which can operate in the mesoscopic photon-number domain, and the self-consistency of our method of analysis allow us to retrieve information about the relative phase of displaced bracket states by simply investigating the statistical properties of the detected-photon numbers. 
\par
Work is still in progress to implement quasi optimal phase-estimation strategies based on the capability of our detectors to discriminate photons, by investigating the performance and limits of the direct inversion of the Fano factor $F(\phi)$ with respect to the Bayesian analysis of the output statistics.\cite{phase-est}

\section*{Acknowledgments}
SO would like to thank Matteo Bina for useful and stimulating discussions. This work has been supported by MIUR (FIRB ``LiCHIS'' - RBFR10YQ3H).

\end{document}